\documentstyle[epsfig]{qcdparis}
\begin{document}
\pagestyle{plain}
\title{Interplay between polarized DIS and RHIC spin physics}

\author{A. Sch\"afer}

\affil{Institute for Theoretical Physics\\
University Frankfurt \\
D-60054 Frankfurt}

\abstract{The complementarity of polarized DIS experiments and
polarized Proton-Proton experiments is illustrated for two examples. It is
shown how
the twist-3 part of the second moment of $g_2(x,Q^2)$
and the single-spin photon asymmetry are connected and it is discussed
how the polarized gluon distribution can be obtained from the measurement of
direct gamma asymmetries.}

\resume{~}

\twocolumn[\maketitle]
\fnm{7}{Talk given in the
 session `Spin Effects in Lepton-Nucleon Reactions'
at the Workshop on Deep Inelastic scattering and QCD,
Paris, April 1995}
\section{Introduction}
In the future the information gained from polarized DIS-experiments
will be suplemented by additional information from polarized proton-proton
collisions planned to be investigated at RHIC starting in the year 2000
\cite{RSC}.
There are several very interesting experiments to be done with polarized
protons \cite{PA,promptg}. The most obvious ones are those allowing to
determine the polarized
gluon structure function $\Delta G(x)$. Several experiments along these lines
have been proposed, for brevity we concentrate here only on one of these,
namely
the detection of spin-asymmetries in direct photon production for the collison
of longitudinally polarized protons \cite{SG1}.
The knowledge of $\Delta G(x)$ is crucial for the interpretation of polarized
DIS because the polarized structure functions measured in DIS cannot be
separated from a possible anomalous gluon contribution. The form of this
contribution, especially its $x$-dependence  is much debated. For the
perturbative contribution one can derive it from standard QCD
\cite{anoma}.
\begin{equation}
\Delta g_1^p(x) = \langle {e^2\over 2} \rangle ~\int_x^1~
A \left( {x\over z}, {Q^2\over \mu_{\rm fac}^2}\right)
\Delta G(z,\mu_{\rm fac}) {{\rm d}z
\over z}
\end{equation}
Here $\mu_{\rm fac}$ is the factorization scale and $A$ is a known splitting
function.
However, this contribution depends
strongly  on the infrared regulators used, which signals
that the distinction between a quark and a gluon part is ill-defined.
Furthermore one would expect in addition strongly  non-perturbative
contributions.
Although this discussion is in no way settled everybody agrees that the
correction term is negligable unless $|\Delta G|$ is of order 1. To decide
whether  this is the case one can analyse not totally inclusive
reactions in DIS like those in which a positive or negative
pion is produced \cite{AS1}. This will be tried by the HERMES
experiment but it looks very difficult for the precission
needed. The other possibility is to investigate
polarized hadron-hadron collisions. We have analysed the latter
possibility using a specialized Monte-Carlo programm called SPHINX \cite{SG2}
based on PYTHIA \cite{Sjo92}.

\section{Prompt-$\gamma$-Production\label{pgp}}
We
investigated the two leading processes
(i.e. first order in $\alpha_s$)
for prompt-$\gamma$-production,
namely the Compton process (\fref{fig4}) and the
annihilation process (fref{fig5})  and determined their
contribution at RHIC energies to the cross section for
different parton parametrisations.
\ffig{disfig4.eps}{3cm}{\em The Compton graph}{fig4}
\ffig{disfig5.eps}{3cm}{\em The annihilation graph}{fig5}
The relevant spin-difference cross-section is
\begin{eqnarray}
&&E_\gamma\frac{{\rm d}\Delta\sigma_{pp\rightarrow \gamma
X}}{{\rm d}^3p_\gamma}(s,x_F,p_\perp)\label{gammapolall}\nonumber \\
&=&
\sum_{ab}\int \ {\rm d} x_a  \ {\rm d} x_b \ \Delta P_a(x_a,Q^2) \Delta
P_b(x_b,Q^2) \\
&&E_\gamma\frac{{\rm d}\Delta\hat{\sigma}_{ab\rightarrow\gamma
X}}{{\rm d}^3p_\gamma}(\hat{s},x_F,p_\perp).\nonumber
\end{eqnarray}
Here the sum is over all partonic subprocesses which contributes to the
reaction $pp\rightarrow \gamma X$.
$\Delta P_a$
and $\Delta P_b$ denotes the polarised parton distribution functions.
The latter are the difference
between the probability to hit partons of the same
helicity as the hadron and those of
opposite helicity.
For our simulation we used two parametrisations
for parton densities with large gluon
polarisation by
Altarelli\&Stirling \cite{altsti} and by Ross\&Roberts (set D)
\cite{rosrob}
and one parametrisation with a small gluon polarisation by
Ross\&Roberts (set A).
\\
For large gluon polarisation the Compton process is the by far dominant one
and one can safely neglect the contribution of the annihilation
process in (\ref{gammapolall}). In this case the
prompt-$\gamma$-production becomes proportional to $\Delta G$ and is
thus a clean probe for the gluon polarisation.
However, in a scenario with a large sea contribution to the spin of
the proton and a gluon polarisation only due to Altarelli-Parisi
evolution, as described by the parametrisation Ross\&Roberts set A
\cite{rosrob},
the annihilation process becomes the major contribution.
For the unpolarised parton distributions we haven chosen the
parametrisation of Gl\"uck, Reya, and Vogt \cite{grv}.
The matrix elements are implemented in leading order only. However,
due to the initial and final state shower algorithm some features of
higher order effects are incorporated as well \cite{pyman}. Also the
polarisation effects are traced in the initial state shower.
For the simulations the polarised initial state shower and the final
state shower were switched on. To avoid infrared divergences
the hard interaction cross section must be supplemented by a lower
cut off for the transverse momentum $p_\perp$. We chose
$p_\perp\ge4\ {\rm GeV}$.

Some results of these simulations are shown in  \fref{fig1},
\fref{fig2} and
\fref{fig3}.
In
 \fref{fig1} the Lorentz-invariant cross section for
prompt-$\gamma$-production as a function of $p_\perp$ at $x_F\approx 0$
is displayed for the Compton process (upper plot) and the annihilation
process (lower plot).
\ffig{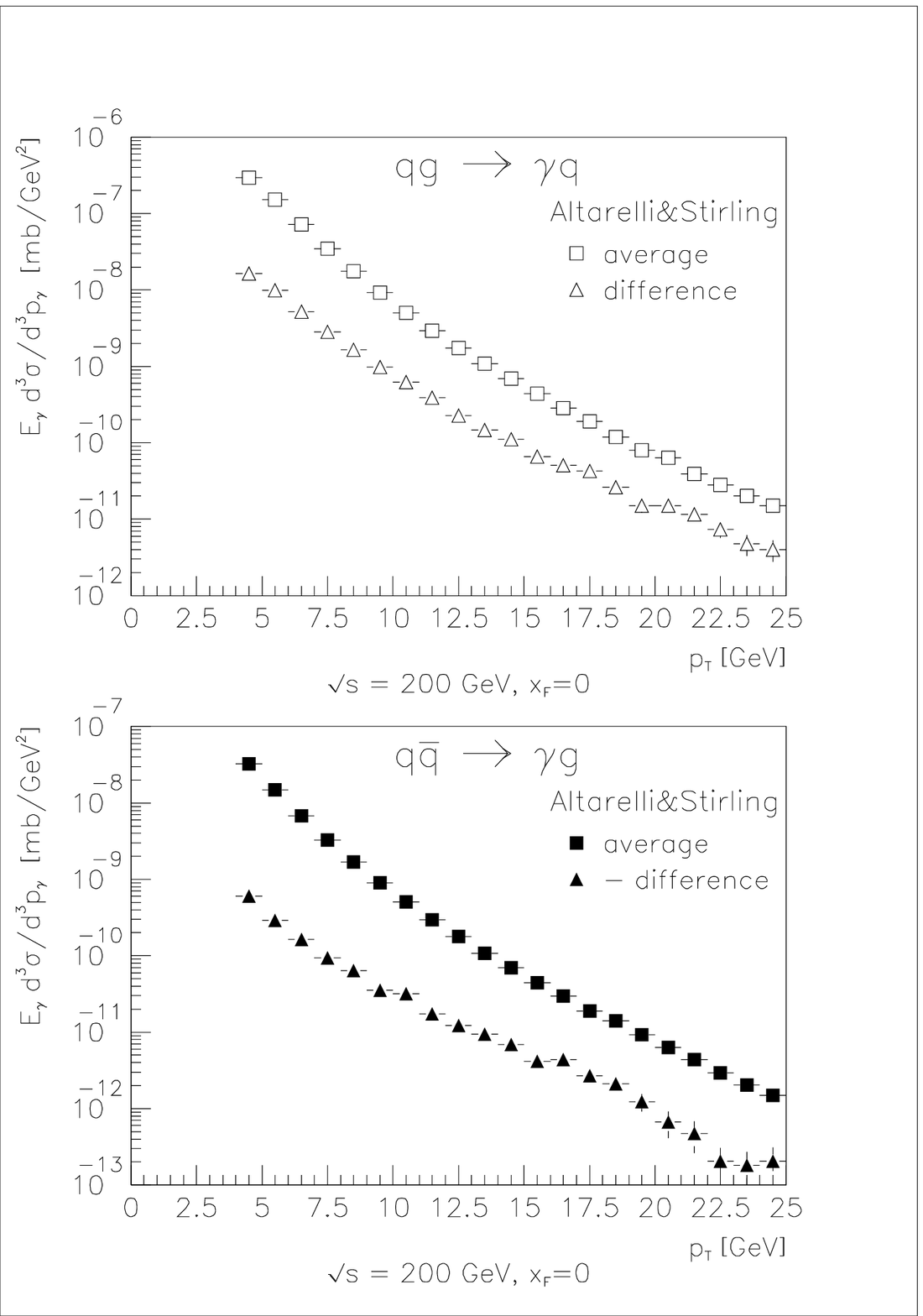}{12.5cm}{\em Spin-average and spin-difference cross section
for prompt $\gamma$
 production}{fig1}

$x_F$ is here the longitudinal momentum fraction of the photon
defined by $x_F=2p_z^{\gamma}/\sqrt{s}$.
In both cases the spin averaged cross section
(squares) and the cross section for the spin difference (triangles)
are shown.
The error bars reflect the MC error.
A typical RHIC run has
320 pb$^{-1}$, such that the $10^7$ events we generated for each spin
combination
(with $p_{\perp}\ge 4$ GeV)
corresponds to an integrated cross section of $3
\cdot 10^{-5}$ mb for the spin averaged case respectively to a
differential cross-section of roughly $3\cdot 10^{-5}~ {\rm mb}/
(4\pi\cdot 4 {\rm
 ~GeV})=6 \cdot 10^{-7}$ mb/GeV in the 4 GeV bin. This implies that the MC
error is roughly comparable to the expected experimental error for
the prompt-$\gamma$'s and substantially larger than the
anticipated experimental errors for gammas from $\pi^0$ decays.

The fact that the spin-differences are clearly non-zero shows that with
this experiment one can indeed measure $\Delta G$.

\subsection{Background considerations}
High-$p_\perp$ $\gamma$'s are not only produced in the direct
processes but at a far larger
rate due to bremsstrahlung and in particular in meson decays. This
background has to be separated from the direct photons very accurately
in order to do not contaminate the signal substantially.
This issue has many aspects we only want to illustrate them by addressing
two points:\\
1.) The detector has to fullfill certain requirements in order to make the
experiments possible, and\\
2.) The direct photons from a so-called background, namely $\pi^0$-decay
offer a very promissing signal. \\
An inportant problem is e.g. whether the detector can at all
identify whether a photon comes from a $\pi^0$-dcay or not. The most important
issue is whether the two photon from pion decay can be distinguished, or
whether they end up in the same detector.
The rate of such `fake' $\gamma$'s dependens obviously on
the spatial detector resolution.
The minimal opening angle of a $\gamma$-pair in the
rest frame of the pion is given by:
\begin{eqnarray}
\chi_{\rm min}&=&2\frac{m_\pi}{E_\pi} ~~~~,\label{chimin}
\end{eqnarray}
and the following resolutions of the detector are considered
$\chi^{\rm res}>0.005\ {\rm rad}$,
$0.01\ {\rm rad}$, or
$0.02\ {\rm rad}$.
Defining the fake-$\gamma$-rate $R$ as the ratio between the number of
unresolved pions and the total number of pions
\fref{fig2}. These plots show that the planned PHENIX
resolution of $\chi^{\rm res}>0.01$ is marginally sufficient to keep the
fake-$\gamma$-rate down.
\ffig{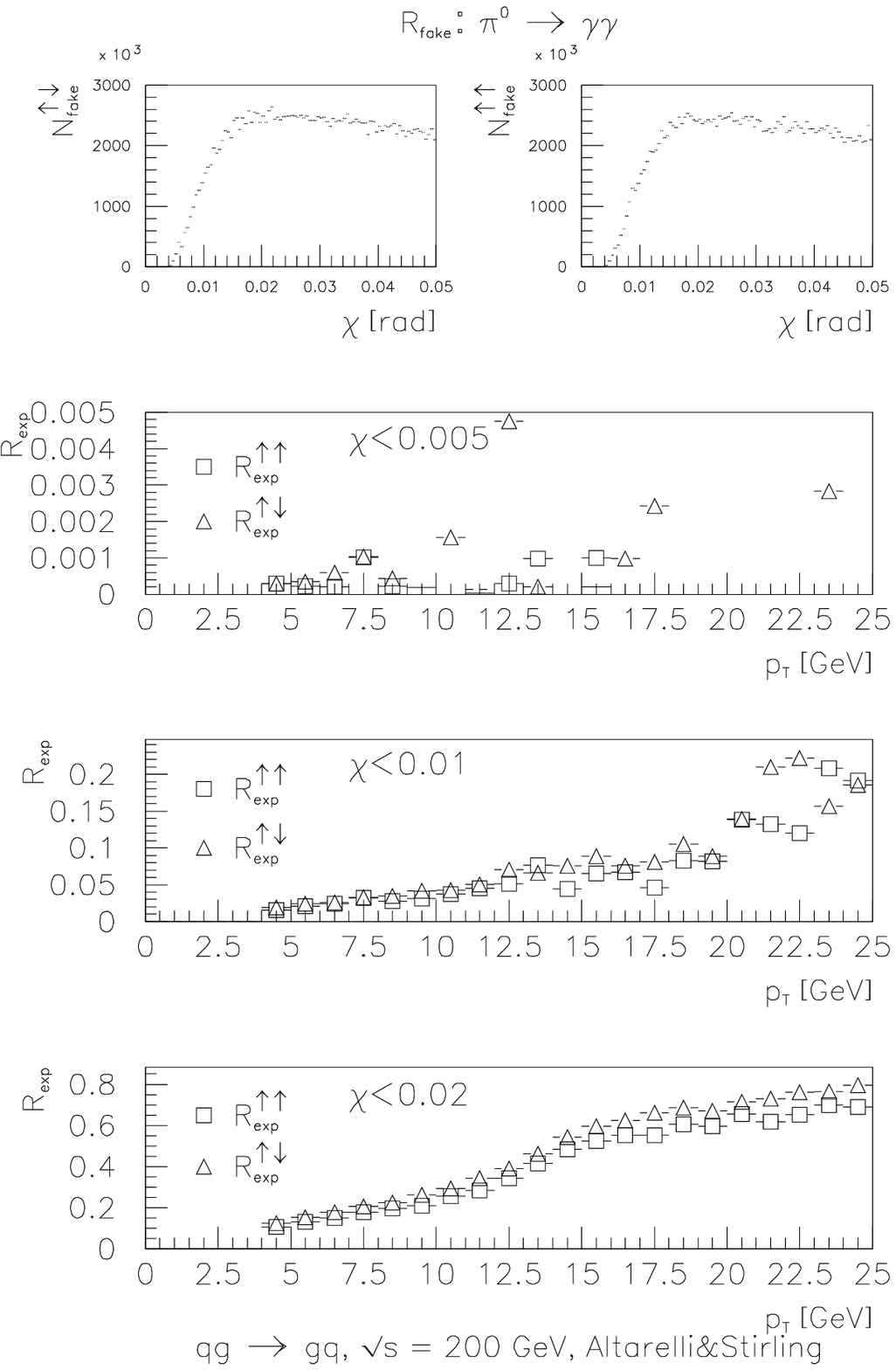}{12.5cm}{\em The background from misidentified $\gamma$'s
from $\pi^0$-decay}{fig2}

Figure 5 shows the yield of pions due to the QCD-Compton process
in the spin averaged case (upper
plot), for the spin-difference (middle) and the resulting asymmetry
(lower plot) as
a function of $p_\perp$ for parametrisations with a large gluon
polarisation. The experimental statistical precission will be much
better than our Monte Carlo errors. Consequently this looks like a very nice
signal indeed.
\ffig{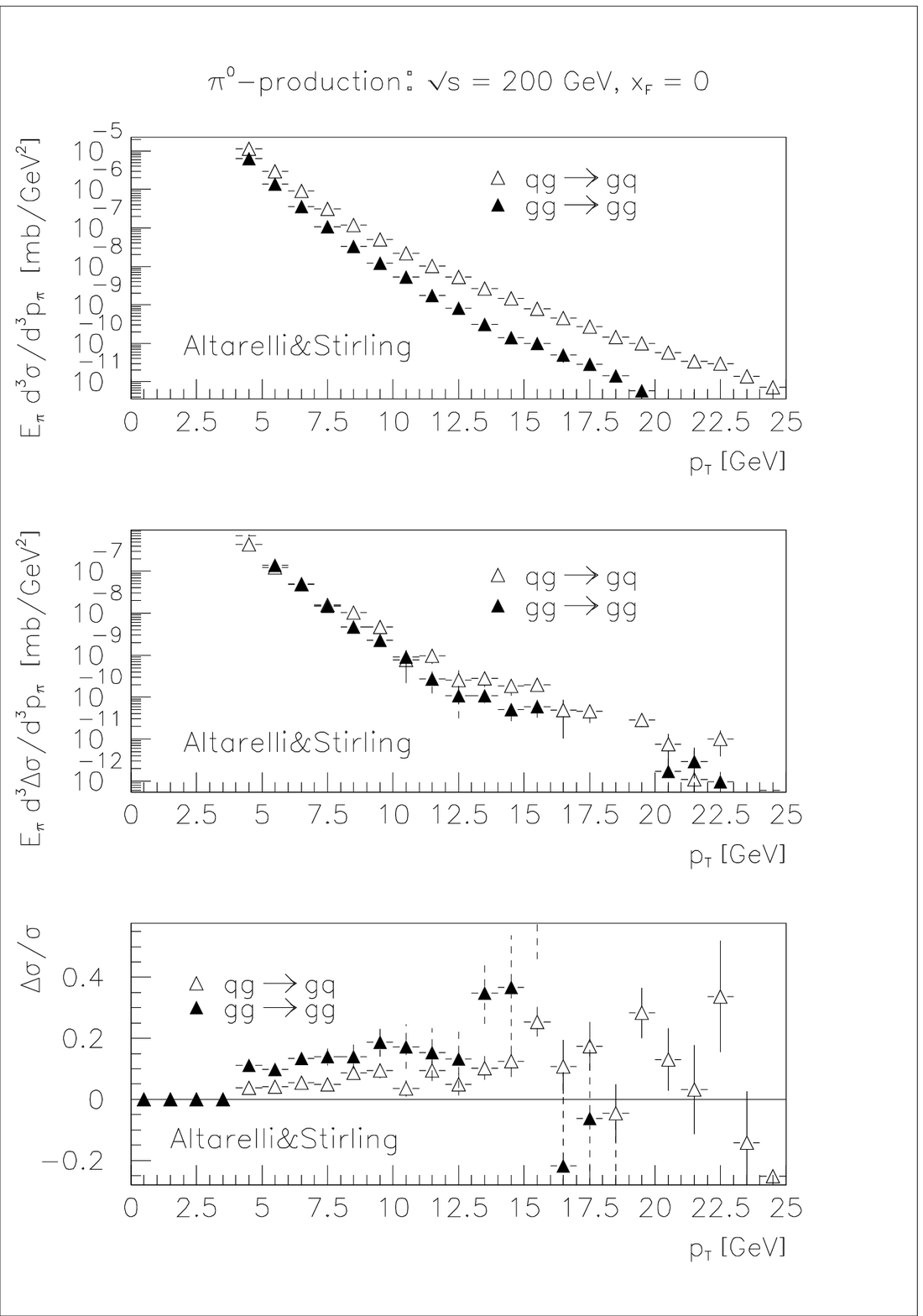}{13cm}{\em The spin-asymmetry in $\pi^0$-production as
signal
for the polarised gluon distribution}{fig3}

\section{$d_2$ and single-spin gamma asymmetries in p+p($\uparrow$)}
By measuring the second moments of the two spin structure functions $g_1(x)$
and $g_2(x)$ one will be able to determine the twist three contribution
$d_2$ which is determined by a specific quark-gluon correlator \cite{theo1}.
\begin{eqnarray}
\int_0^1 x^2 g_1(x,Q^2) = {1\over 2} a_2 + O(M^2/Q^2) \nonumber \\
\int_0^1 x^2 g_2(x,Q^2) = -{1\over 3} a_2 +{1\over 3} d_2+ O(M^2/Q^2)
\end{eqnarray}
\begin{equation}
d_2 \sim \langle P,S| \bar \psi \left[\gamma_{\alpha}g \tilde G_{\beta
\sigma} +\gamma_{\beta}g \tilde G_{\alpha
\sigma}- trace \right]\psi |PS\rangle
\end{equation}
These relations follow from operator product expansion  and are thus
firmly rooted in the framework of QCD.
$d_2$ plays in some way the analogous role to the magnetic field in QED.
In fact the situation is slightly involved, because the QED analogy can
be rather misleading. In QCD the colour-magnetic field and possible
correlations
between transverse momenta and colour-electric fields are of equal importance.
Also the fields must always be coupled to colour-singlets, e.g. by coupling
them to quark and antiquark as for $d_2$. Still $d_2$
should reappear in many different places in spin physics, just as the
magnetic field
does for QED.
One such place is the production of direct photons in the collision of
a transversely polarized nucleon with an unpolarized one. One can argue for
this case that the photon asymmetry $A_{\gamma}$
scales with $d_2$, such that
\begin{equation}
{A_{\gamma}^{\rm proton} \over A_{\gamma}^{\rm neutron}} =
{d_2^{\rm proton} \over d_2^{\rm neutron}} ~~~~~.
\end{equation}
As all four quantities will be measured
(assuming that the necessairy precission for the photon asymmetries
can be reached experimentally, which requires a dedicated effort)
it should be clear within a few years
whether this prediction is fullfilled.

\section{Summary}
We have
demonstrated for two simple examples how fruitful the interplay between
polarized DIS and polarized proton-proton-collisions is. RHIC spin physics
will make a major contribution to the understanding of the nucleon spin
structure.

\begin{center}
{\large\bf Aknowledgements}
\end{center}
It is a pleasure to thank the convenors of the
spin-working group and the organizing committee, especially J. Soffer.


\begin{thebibliography}{99}
\bibitem{RSC} G. Bunce et al., Particle World 3 (1992) 1\\
Joined PHENIX-STAR spin proposal 1993
\bibitem{PA}
Proceedings of the `Polarized Collider Workshop', University Park, PA
1990,\\
edts. J. Collins, S.F. Heppelman, and R.W. Robinett, AIP conf. proc.
No 223, New York 1991
\bibitem{promptg} E.L. Berger and J. Qiu, Phys. Rev. D40 (1989)
778,\\
Phys. Rev. D40 (1989) 3128, and J. Qiu in \cite{PA}\\
C. Bourelly, J. Ph. Gullet, and J. Soffer, Nucl Phys. B 361 (1991) 72
\bibitem{anoma} C.S. Lam and B.-A. Li, Phys. Rev. D25 (1982) 683\\
R.D. Carlitz, J.C. Collins, and A.H. Mueller, Phys. Lett. B214 (1988)
229\\
G. Altarelli and G.G. Ross, Phys. Lett. B212 (1988) 3911\\
J. Ellis, M. Karliner, and C.T. Sachrajda, Phys. Lett. B231 (1989)
497\\
L. Mankiewicz and A. Sch\"afer, Phys. Lett. B242 (1990) 455
R.L. Jaffe and A. Manohar, Nucl. Phys. B337 (1990) 509
\bibitem{AS1} St. G\"ullenstern et al., Phys. Lett. B312 (1993) 166
\bibitem{SG1} St. G\"ullenstern et al., Phys. Rev. D 51 (1995) 619
\bibitem{SG2} St. G\"ullenstern et al., Comp. Phys. Comm. to be published
\bibitem{Sjo92}T. Sj\"ostrand, {\sc PYTHIA} 5.6 and {\sc JETSET} 7.3, Physics
and
Manual,\\
CERN-TH.6488/92 May 1992 (revised Sept. 1992)
\bibitem{altsti} G.~Altarelli and J.~Stirling, Particle World 1 (1989) 40.
\bibitem{rosrob} G.\ G.\ Ross and R.\ G.\ Roberts, Rutherford preprint
RAL-90-062 (1990).
\bibitem{grv} M. Gl\"uck, E. Reya, and A. Vogt, Z. Phys. C48
(1990) 471
\bibitem{pyman} T.~Sj\"ostrand, CERN-TH.6488/92; T.~Sj\"ostrand,
Comp. Phys.
Comm. 39 (1986) 347; T.~Sj\"ostrand and M.~Bengtsson, Comp. Phys.
Comm. 43 (1987) 367.
\bibitem{theo1}
I.I. Balitsky, V.M. Braun and A.V. Kolesnichenko, Phys. Lett. B 242
(1990) 245; B318 (1993) 648(E).\\
E. Stein et al.,
`QCD Sum Rule Calculation of Twist-3 Contribution to Polarised
Nucleon Structure Functions', UFTP preprint 366/1994, HEP-PH-9409212 \\
X. Ji and P. Unrau, Phys. Lett. B333 (1994) 228\\
B. Ehrsperger, L. Mankiewicz, and A. Sch\"afer, Phys. Lett. B323
(1994) 439\\
\bibitem{Ehrns} B. Ehrnsperger et al., Phys. Lett. B321 (1994) 121
\end{thebibliography}
\end{document}